\documentclass[
reprint,
superscriptaddress,
amsmath,
amssymb,
aps,
prl,
floatfix,
]{revtex4-1}

\usepackage{footnote}	
\usepackage{braket}
\usepackage{dsfont}
\usepackage{chemformula}
\usepackage{subfigure}
\usepackage{graphicx}% Include figure files
\usepackage{dcolumn}% Align table columns on decimal point
\usepackage{bm}% bold math
\usepackage{relsize}
\usepackage{slashed}
\usepackage{dirtytalk}
\usepackage{amsmath}
\usepackage{siunitx}
\usepackage{layouts}
\usepackage{xspace}
\usepackage{graphicx,color,hyperref}
\hypersetup{colorlinks=true, linkcolor=blue, citecolor=blue, urlcolor=blue} 

\usepackage{braket}
\usepackage{bm}% bold math
\usepackage{xcolor}
\usepackage{amsmath}
\usepackage{amssymb}
\usepackage{siunitx}
\usepackage{hyperref}% add hypertext capabilities

\newcommand{\om}{\omega}

\newcommand*\dd{\mathop{}\!\mathrm{d}} %differential d%
\renewcommand{\vec}[1]{{\bf #1}}

\newcommand{\kv}{\vec{k}}

\usepackage{multibib}  % Enables multiple bibliographies
\newcites{A}{References for Part A}  % Bibliography for first part
\newcites{B}{References for Part B}  % Bibliography for second part

\begin{document}

%\preprint{APS/123-QED}

\title{Field-driven band asymmetry and non-reciprocal transport in a helimagnet}
% Force line breaks with \\

%%%%%%% List of authors %%%%%%
\author{Darius-Alexandru Deaconu}
\email{darius-alexandru.deaconu@manchester.ac.uk}
\affiliation{Department of Physics and Astronomy, The University of Manchester, Oxford Road, Manchester M13 9PL, United Kingdom}
\author{Aneesh Agarwal}
\affiliation{Theory of Condensed Matter Group, Cavendish Laboratory, University of Cambridge, J. J. Thomson Avenue, Cambridge CB3 0HE, UK}
\author{Rodion Vladimirovich Belosludov}
\affiliation{Institute for Materials Research, Tohoku University, Sendai 980-08577, Japan}
\author{Robert-Jan Slager}
\affiliation{Department of Physics and Astronomy, The University of Manchester, Oxford Road, Manchester M13 9PL, United Kingdom}
\affiliation{Theory of Condensed Matter Group, Cavendish Laboratory, University of Cambridge, J. J. Thomson Avenue, Cambridge CB3 0HE, UK}
\author{Mohammad Saeed Bahramy}
\email{m.saeed.bahramy@manchester.ac.uk}
\affiliation{Department of Physics and Astronomy, The University of Manchester, Oxford Road, Manchester M13 9PL, United Kingdom}

\date{\today}

\begin{abstract}
Helimagnets exhibit noncollinear spin arrangements characterized by a periodic helical modulation, giving rise to emergent chiral properties. These materials have attracted significant interest due to their potential applications in spintronics, particularly for robust information storage and the realization of topological spin textures such as skyrmions. In this work, we focus on Yoshimori-type helimagnets, where competing exchange interactions mediated by conduction electrons stabilize helical spin structures without requiring Dzyaloshinskii-Moriya interaction. We introduce a minimal model describing the electronic structure of a one-dimensional helimagnet in the presence of an external magnetic field and investigate its impact on non-reciprocal transport. We demonstrate how band asymmetry emerges in the conical phase induced by the external field, leading to a nonzero second-order electronic conductivity and injection photoconductivity. Our results provide insight into the interplay between the real space magnetic texture and electronic properties, paving the way for future studies on chirality-driven transport phenomena in centrosymmetric helimagnets.
\end{abstract}

\maketitle
\textit{Introduction}\textemdash Chirality is a fundamental concept in physics, chemistry, and biology, which describes the property of an object or system that cannot be superimposed on its mirror image. This intrinsic handedness emerges across scales, from the molecular arrangements of organic compounds to the macroscopic structures of certain crystals and biological systems \citeA{wagniere2007chirality}. In condensed matter physics, chirality plays a central role in phenomena such as nonreciprocal charge transport, topological textures, and magnetoelectric effects \citeA{bousquet2024structural}. It can arise as a consequence of the crystal structure of a material, depending on the space group describing the symmetry \citeA{felser2022topologychirality}. In this regard, we note a recent interaction in using (magnetic-spin) space group input and topological structures~\citeA{Watanabe2018, Bernevig2022, kruthoff2017, Xiao2024, Bouhon2021, Yi2024, Corticelli2022}. However, chirality can arise not only from the crystal structure but also from the magnetic order within the material, as seen in the case of helimagnets.

In the context of helimagnetism, chirality manifests through spiral spin configurations where neighbouring magnetic moments arrange themselves at a fixed angle, forming a helical pattern \citeA{yoshimori1959new}. Helimagnetic systems provide a rich playground for exploring the physics of chirality, as their unique spin textures can host exotic quasiparticles like skyrmions \citeA{fert2017magnetic}, which can be interpreted as a superposition of multiple helimagnetic structures \citeA{muhlbawer2009skyrmions}, and contribute to emergent properties such as topological Hall effects \citeA{nagaosa2013topological}. In terms of device applications, the two distinct enantiomers offer a promising platform for use as binary states, enabling robust information storage \citeA{tokura2010multiferroics, masuda2024room}. Therefore, understanding the origins and consequences of chirality in helimagnetic materials is essential for developing advanced spintronic devices and uncovering new paradigms of magnetically-driven phenomena.

\begin{figure}[t]
    \centering
    \includegraphics[width=0.98\linewidth]{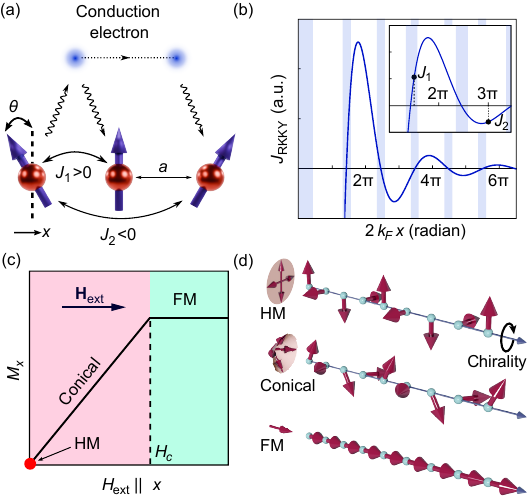}
    \caption{RKKY interaction inducing helimagnetic ordering. (a) Coupling mechanism required to obtain a helimagnetic configuration stabilised by competing exchange interactions resulting from mediated exchange (RKKY interaction). (b) A plot of the RKKY coupling proportionality to a function $F(2k_{\mathrm{F}}x)$, where $x$ is the distance and $k_{\mathrm{F}}$ is the Fermi wavevector. The shaded regions indicate where $J_1$ and the corresponding $J_2$ meet the conditions for stabilizing helimagnetic order, as illustrated in the inset. (c) Sketch of the magnetisation along the chain (x-direction) as a function of an external magnetic field applied perpendicular to the helical plane. (d) Magnetic configurations of the spin spiral in an external field, from top to bottom in the order of increasing field: helimagnetic phase at $H_{\mathrm{ext}=0}$, conical phase for $0 < H_{\mathrm{ext}=0} < H_c$, and field-forced ferromagnetic phase for $H_{\mathrm{ext}=0} \geq H_c$.}
    \label{fig:figure1}
\end{figure}

The study of helimagnetic structures began more than half a century ago and has since developed into a rich field of research. It was found that such states are often stabilized by various mechanisms, including the presence of competing exchange interactions or antisymmetric spin couplings in systems lacking inversion symmetry. The former mechanism was introduced by Yoshimori \citeA{yoshimori1959new}, Kaplan \citeA{kaplan1959classical}, and Villain \citeA{villain1959structure}, who identified the magnetic structure of \mbox{Mn$\mathrm{O}_2$} \citeA{erickson1953neutron} as helimagnetic. The latter was introduced by Dzyaloshinskii \citeA{dzyaloshinskii1964theory}, who identified that the relativistic spin-orbit coupling in a non-centrosymmetric crystal can give rise to an antisymmetric interaction, known as Dzyaloshinskii-Moriya (DM) \citeA{dzialoshinski1958thermodynamic, moriya1960anisotropic}. Typically, this situation occurs in crystals with chiral space groups, such that the symmetry is restricted to proper rotations~\citeA{Watanabe2018, Bernevig2022, kruthoff2017, Xiao2024, Bouhon2021, Yi2024, Corticelli2022}. Naturally, this interaction favours spin canting, and its competition with isotropic ferromagnetic exchange can stabilise helimagnetic structures. An extensively studied material candidate for a DM-type helimagnet is the intercalated transition-metal dichalcogenide $\mathrm{Cr}_{1/3}\mathrm{NbS}_2$, in which the formation of chiral magnetic soliton lattice has been observed \citeA{moriya1982evidence, cao2020overview, kousaka2022emergence, togawa2012chiral}. Another candidate is the chiral magnet $\mathrm{Co}_8\mathrm{Zn}_9\mathrm{Mn}_3$ where nonreciprocal resistivity was measured at room temperature, and for which a theoretical model that identifies the mechanisms responsible was proposed recently \citeA{nakamura2024nonreciprocal}.

The manipulation of chirality poses a challenge in this case because the degeneracy between left- and right-handed spiral structures is already lifted at the Hamiltonian level \citeA{togawa2016symmetry}. The spin-rotation direction is fixed by the noncentrosymmetric crystal structure, making it difficult to be reversed by external fields. On the other hand, the control of chirality using external fields has been achieved experimentally in the centrosymmetric compounds MnP, $\text{MnAu}_2$ and \mbox{$\alpha\text{-EuP}_3$} \citeA{jiang2020electric, masuda2024room, mayo2025band}. In the case of \mbox{$\alpha\text{-EuP}_3$}, the non-reciprocal charge transport has been measured, and its microscopic origin was related to the asymmetry of the band structure \citeA{mayo2025band}. Because the systems mentioned above have a centrosymmetric crystal structure, the DM interaction is absent. Motivated by our results obtained for \mbox{$\alpha\text{-EuP}_3$}, the focus of this paper is to introduce a simple model that describes the electronic structure of a Yoshimori-type helimagnet in an external magnetic field. The model captures the band asymmetry responsible for the non-reciprocal transport.

\textit{Yoshimori helimagnetism}\textemdash An essential feature of the exchange interaction required for the Yoshimori-type helimagnet is that it extends beyond the nearest neighbour. If we denote the interaction between two neighbouring spins by $J_1$, and between the next-nearest neighbours by $J_2$, one can obtain a helical arrangement of spins in the lowest energy state when $|J_1|<|4J_2|$, and $J_2<0$ \citeA{nagamiya1968helical}. Intuitively, a positive (negative) value of $J_1$ means that the spins on neighbouring sites want to become parallel (antiparallel) to minimise energy, while a negative $J_2$ implies that the next nearest neighbours need to be antiparallel to do so. Therefore, a compromise between the two situations results in neighbouring spins becoming canted, which gives an overall helical arrangement. Generally, the resulting spin structure is incommensurate with the period of the lattice, since the angle $\theta$ between two neighbouring spins is determined only by the exchange coefficients. The situation described above is shown in Figure \ref{fig:figure1} (a), where the competing exchange interactions arise from a mediated exchange through conduction electrons, known as Ruderman-Kittel-Kasuya-Yosida (RKKY) interaction \citeA{ruderman1954indirect, kasuya1956theory, yosida1957magnetic}. This commonly occurs in the case of transition metals and rare earth elements, where the magnetic moments of strongly localised $d$- and $f$-electrons, respectively, interact with each other via a mediated exchange through $s$-electrons or $p$-electrons \citeA{yosida1996theory}. The RKKY interaction fulfils enough conditions to stabilise spiral spin arrangements since it is long-ranged, and also oscillatory. By approximating the conduction electrons as free, the behaviour of the exchange coupling is shown in Figure \ref{fig:figure1} (b). It can be seen that sign changes occur as the distance increases. The highlighted areas indicate the regions in which the conditions for a stable helimagnet are satisfied. An example of such a situation is shown in the inset of Figure \ref{fig:figure1} (b), where the distance $a$ between nearest neighbours is such that $2k_{\mathrm{F}}a = 3\pi/2$, which gives a positive $J_1$, a negative $J_2$, and $|4J_2|>|J_1|$ is satisfied. 

The periodic modulation of the spin texture in helimagnets is characterised by the propagation vector ($\bm q$-vector), which describes how the spin vector rotates as the position advances along $\bm q$. Generally, the $\bm q$-vector is not perpendicular to the rotation plane, and the relative orientation between the screw axis and the rotation plane is often dictated by magnetic anisotropy. An easy-plane anisotropy favours spin rotation within that plane, while $\bm q$ aligns based on anisotropies in the exchange interaction \citeA{nagamiya1968helical}. Therefore, when the system is placed in an external magnetic field, the behavior of the spin structure is expected to depend on the orientation of the field relative to the spin rotation plane. Assuming that the external field is applied perpendicular to the rotation plane, the phase diagram of the helimagnet is shown in Figure \ref{fig:figure1} (c). 

The evolution of the spin structure with the increasing external field is displayed in Figure \ref{fig:figure1} (d). When no external field is applied, the system is in the helimagnetic (HM) phase. As the magnetic field increases, the spins are tilting towards the direction of the field forming a conical phase. The system becomes ferromagnetic when the field increases above the critical threshold, $H_c$.

\textit{Model Hamiltonian}\textemdash In order to capture the evolution of the electronic structure and the signatures of each phase in transport properties, we introduce a simplified model for a one-dimensional chain of atoms with helical spin order. In this simplified model, we consider \mbox{$\bm q  = (\pi/4,0,0)$}, such that the magnetic moments on neighbouring sites are rotated around the x-axis by $90^\circ$.

\begin{figure}[t]
    \centering
    \includegraphics[width=0.98\linewidth]{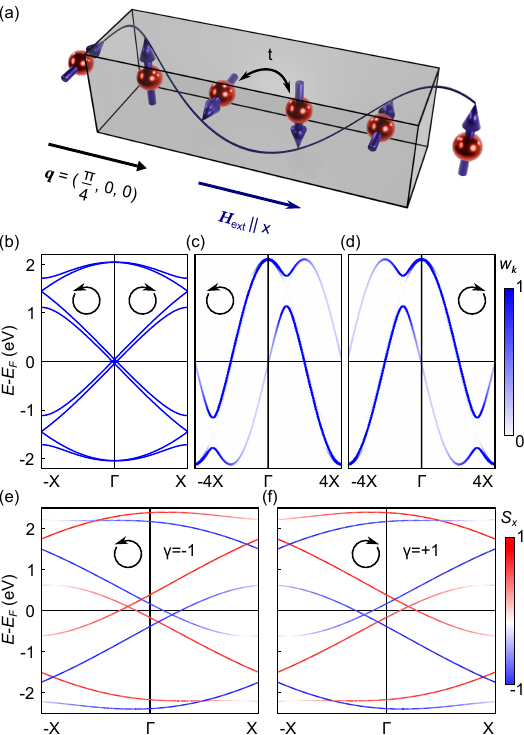}
    \caption{Chirality dependence of electronic band structure in a one-dimensional helimagnet. (a) One-dimensional chain of atoms with helimagnetic order for which the $\bm q-$vector is along the chain. The electron hopping amplitude between nearest-neighbour sites is $t$. (b) Band structure for $J=0.3$ in the helical phase, which looks identical for both possible chiralities of the spiral. (c), (d) Unfolded band structures for the cases in which the helix has anti-clockwise and clockwise chirality, respectively. The colour indicates the spectral weight. (e), (f) Spin projected band structure for $\gamma=-1$ and $\gamma=+1$, respectively, in the conical phase with tilt angle $\alpha = 10^{\circ}$ and $J=0.8\,\mathrm{eV}$. }
    \label{fig:figure2}
\end{figure}

The Hamiltonian for the one-dimensional chain of atoms shown in Figure \ref{fig:figure2} (a) can be written as
\begin{equation}
    H = \sum_{<i,j>} t c^\dagger_ic_j + h.c. + J\sum_{i}  c^\dagger_i \bm M_i \cdot \bm \sigma c_i,
\end{equation}
where the first sum is taken over the nearest-neighbours, $t$ is the electron hopping amplitude between site $i$ and site $j$, and $J$ is the coupling between $s$ orbitals and localised $d$ or $f$ orbitals. $\bm M_i$ is the local magnetization at site $i$ and $\bm \sigma = (\sigma_x,\sigma_y,\sigma_z)$ is the vector of Pauli matrices. If an external magnetic field, $\bm H_{\mathrm{ext}}$, is applied along the direction of the chain, perpendicular to the plane of the spiral, each magnetic moment will be tilted by an angle $\alpha$ away from this plane. Hence, the magnetisation on each site in the unit cell becomes
\begin{equation}
    \bm M_i  = M \Big(\sin \alpha, \gamma \sin\big(\frac{\pi}{2}i\big) \cos \alpha, \cos \big(\frac{\pi}{2}i\big) \cos \alpha \Big), 
    \label{eq:magnetisation_site}
\end{equation}
where $i = 0,1,2,3$ and $M$ is the magnitude of the local magnetic moment. Unless explicitly specified, the values used in our calculations for the parameters are $t=1\,\mathrm{eV}$, $J=0.6\,\mathrm{eV}$ and $M=1$. Recently, a similar Hamiltonian to the one for $\alpha=0$ has been introduced in the context of unconventional $p$-wave magnets, where a minimal tight-binding model was developed to capture key features of p-wave spin-polarized bands protected by a composite time-reversal and translation symmetry \citeA{linder2024minimalmodels}.

In the case in which there is no external field, the magnetizations given in Equation \ref{eq:magnetisation_site} would describe a helical arrangement of spins, such that $\bm q$ is normal to the plane of rotation, as shown in Figure \ref{fig:figure2} (a). The band structure resulting from this arrangement is shown in Figure \ref{fig:figure2} (b), and is independent of the chirality of the spiral since the two chiral states are degenerate. However, if the band structure is unfolded to the primitive unit cell, which contains a single atom, then the resulting band structures are different for the two chiralities. These are shown in Figures \ref{fig:figure2} (c) and (d) for the two opposite chiralities, respectively. The plots are colored according to the spectral weights calculated using the method introduced in \citeA{deretzis2014role}. It can be seen that the band structure resembles the $\cos k_x$ dispersion expected for a single atom, with the additional gaps resulting from the interaction with the local magnetization. To highlight this resemblance, a smaller value of $J=0.3\,\mathrm{eV}$ has been chosen to generate these plots, since the size of the gaps increases with increasing $J$. In contrast to the helimagnetic phase, the band structure asymmetry in the conical phase depends on the chirality of the spin spiral. This dependence is shown in Figure \ref{fig:figure2} (e) and (f), for $\gamma=-1$ and $\gamma=+1$, respectively.

\begin{figure}[t]
    \centering
    \includegraphics[width=0.98\linewidth]{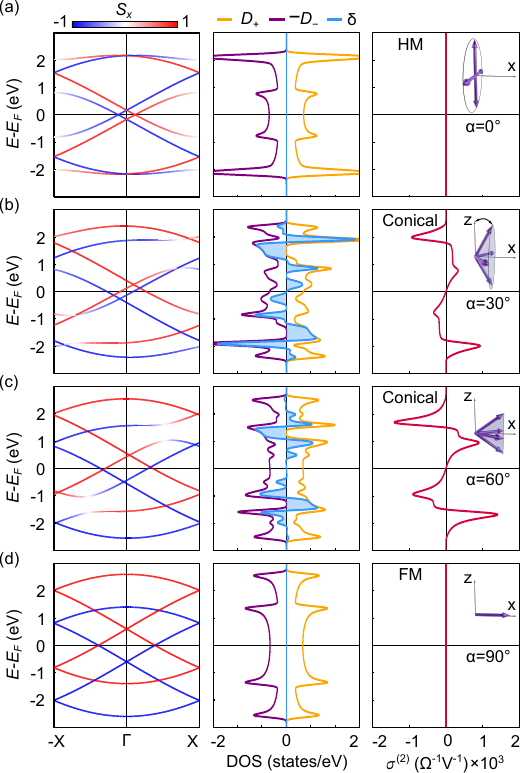}
    \caption{Evolution of the electronic structure in an external magnetic field. (a) Helimagnetic phase, showing (from left to right) the spin-projected band structure, density of states for $k_x\leq0$ (purple, with a negative sign) and for $k_x\geq0$ (orange), and the difference between the two (light blue). The last panel on the right is the second-order conductivity, with the inset showing the arrangement of spins translated to the rotation plane. (b), (c) and (d) show the same information as (a) for the conical phase with $\alpha=30^\circ$, $\alpha=60^\circ$ and ferromagnetic, respectively, where $\alpha$ is the tilt angle of the spins with respect to the $z$-axis.}
    \label{fig:figure3}
\end{figure}

\textit{Evolution under external field}\textemdash To understand the consequences of the band asymmetry on transport properties, we compute the electronic conductivity up to second order using Boltzmann transport formalism under the constant relaxation time approximation. For a one-dimensional system, it is straightforward to show that the $n$-th order of the electronic conductivity at temperature T and chemical potential $\mu$ is given by
\begin{equation}
    \sigma^{(n)}(\mu,T)=-e\!\!\int_k\sum_m v_m(k)\Big(\frac{e\tau}{\hbar}\frac{\partial}{\partial k}\Big)^{n}\!f\big(E_m,\mu,T\big),
    \label{eq:nth_conductivity}
\end{equation}
where the integral is over the entire Brillouin zone, $m$ runs over all bands, $e$ is the elementary charge, $\tau$ is the constant relaxation time, $v_n(k)$ is the velocity of band $m$ at $k$, and $f\big(E_m,\mu,T\big)$ is the Fermi-Dirac distribution function for a state with energy $E_m$. In our calculations, the constant relaxation time is assumed to be $\tau=10\mathrm{ps}$.

\begin{figure}[t]
    \centering
    \includegraphics[width=0.98\linewidth]{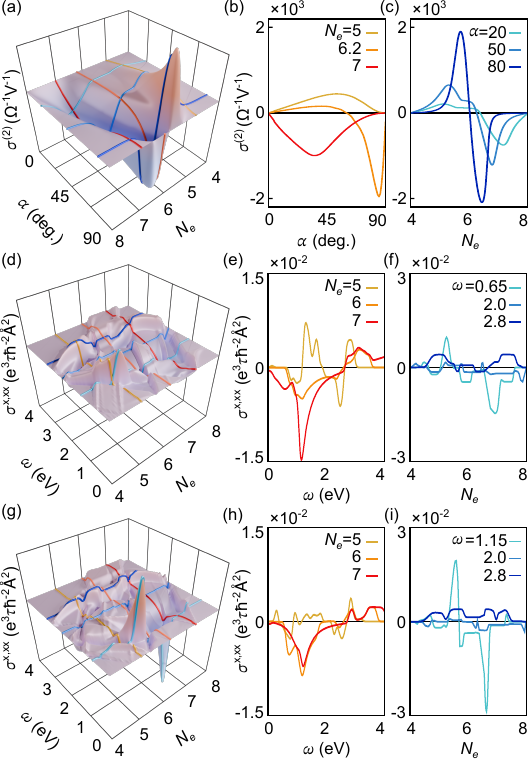}
    \caption{Evolution of second-order conductivity and injection photocoductivity in an external magnetic field. (a) Second-order electronic conductivity evolution with the tilt angle $\alpha$ and for all carrier densities. (b), (c) Cuts through the surface plot in (a) for fixed carrier densities and tilt angles in the conical phase, respectively. (d) Injection photoconductivity dependence on the photon frequency and carrier density in the conical phase with $\alpha=30^\circ$.   (e), (f) Cuts through the surface plot in (d) at fixed values of carrier density and photon frequencies, respectively. (g), (h), (i) the equivalent of (d), (e), and (f), respectively, for $\alpha=60^\circ$.}
    \label{fig:figure4}
\end{figure}

The evolution of the electronic structure when an external magnetic field is applied perpendicular to the spin rotation plane is shown in Figure \ref{fig:figure3}. The electronic band structure in the initial helimagnetic phase is symmetric under the transformation \mbox{$k_x \rightarrow -k_x$}, and the density of states remains identical for the two halves of the Brillouin zone (BZ) corresponding to $k_x \leq 0$ and $k_x \geq 0$. Therefore, in this phase, there is no imbalance in the density of states, and the second-order conductivity vanishes for any chemical potential. This is shown in panel (a) of Figure \ref{fig:figure3}. By applying an external field, the band structure becomes asymmetrical. As a consequence, the difference in the density of states between the two halves of the BZ is non-zero, and the second-order conductivity is non-vanishing in the conical phase. The conical phase is shown in Figure \ref{fig:figure3} (b) and (c) for a tilting angle of \mbox{$\alpha = 30^\circ$} and \mbox{$\alpha = 60^\circ$}, respectively. The symmetry of the bands is restored in the field-forced FM phase, as shown in Figure \ref{fig:figure3} (d), and both the difference in density of states and the second-order electrical conductivity vanish.
A more detailed view of the evolution of second-order conductivity across the conical phase for all possible carrier densities is presented in Figure \ref{fig:figure4} (a). The displayed range of carrier densities corresponds to $\mu \geq 0$, since $\sigma^{(2)}(\mu,T) = -\sigma^{(2)}(-\mu,T)$. The behaviour within the conical phase can be significantly altered by adjusting the carrier density. Figure \ref{fig:figure4} (b) illustrates three specific cases at fixed densities, where the conductivity remains either strictly positive, strictly negative, or changes sign throughout the evolution. The dependence on carrier density is further emphasized in Figure \ref{fig:figure4} (c).

\textit{Optical responses}\textemdash A fundamental question concerns whether the discussed phases can be diagnosed. To that end, we consider optical excitation-induced DC (second-order) responses of the system, comprising shift currents (arising from the positional shift of the Wannier centers upon excitation) and injection currents (arising from differing band velocities of the excited and ground states)~\citeA{Sipe1993, Sipe2000}. Concretely, we use the quantum geometrical framework of optical responses introduced in \citeA{Ahn2021,bouhon2023quantum}, which can phrase these observables in terms of multiband Berry connections $\textbf{A}_{nm} = i \bra{u_{n,\textbf{k}}}\nabla_{\textbf{k}}\ket{u_{m,\textbf{k}}}$.  The shift and injection photoconductivities in response to linearly polarized light can be written as 
% One-column version of this:
\begin{align}
    \sigma^{a,aa}_s (\omega) &= -\frac{2\pi e^3}{\hbar^2} \sum_{m,n} \int_{\text{BZ}} \frac{\text{d} k}{2\pi} \delta(\omega - \omega_{mn}) \nonumber  \\
    & \hspace{2.cm} \times f_{nm,\textbf{k}} ~|A^{a}_{mn}|^2 ~R^{a,a}_{mn} ,\\
    \sigma^{a,aa}_i (\omega) &= -\frac{\pi e^3 \tau}{\hbar^2} \sum_{m,n} \int_{\text{BZ}} \frac{\text{d} k}{2\pi} \delta(\omega - \omega_{mn}) \nonumber  \\
    & \hspace{2.cm} \times  f_{nm,\textbf{k}} |A^{a}_{mn}|^2 ~\partial_a \omega_{mn},
\end{align} 
\noindent respectively. Here, $f_{nm,\textbf{k}} = f_{n,\textbf{k}} - f_{m,\textbf{k}}$ is the difference between the Fermi-Dirac distribution functions of the bands $n$ and $m$, $\omega_{mn} = (E_m-E_n)/\hbar$ is the frequency of the optical transition, and $\tau$ is a relaxation time for the decay of the photoexcited particle. $\textbf{R}^a_{mn} = \textbf{A}_{mm} - \textbf{A}_{nn} - i\nabla_{\textbf{k}} \text{Arg}~ (A_{mn}^a)$ is the 'shift vector' that encapsulates the shift in the Wannier center of the photoexcited particle. 

While the shift vanishes identically at all frequencies for all three phases under consideration, we find that the injection currents are nonzero for the conical phase for nonzero chemical potential, see Fig. \ref{fig:figure4}. This injection behavior thus presents a method for the experimental diagnosis of the conical phase, which is the only phase exhibiting non-zero conductivity. However, it is important to note that this diagnosis is only valid at nonzero chemical potentials since the injection current, being odd in $\mu$, becomes zero when $\mu=0$; therefore, a different test is required under such a condition. 

The behavior of the shift and injection photoconductivities can also be motivated from the symmetries of the model. Since all three phases have chiral symmetry (which can be decomposed into a particle-hole transformation followed by a time reversal \citeA{Ryu_2010}), the shift must necessarily be zero, while no such restriction applies for the injection. This follows from the observation that under a particle-hole transformation, $\sigma^{a,aa}_{s/i} \rightarrow -\sigma^{a,aa}_{s/i}$, whereas under time reversal, $\sigma^{a,aa}_{s/i} \rightarrow \pm \sigma^{a,aa}_{s/i}$ \citeA{Ahn2020}; hence, under the combined action of these transformations, $\sigma^{a,aa}_{s/i} \rightarrow \mp \sigma^{a,aa}_{s/i}$, therefore necessitating a zero-shift response. Additionally, the extra T or P symmetries in the helical or FM phases further make the injection similarly zero~\citeA{Ahn2020} (whereas no such argument can be made for the conical phase). Such a generalized symmetry argument for the behavior of the injection response further highlights the generality of our proposal for the experimental diagnosis of the conical phase using the vanishing (or none thereof) of the injection photoconductivity.  

\textit{Conclusion}\textemdash In this study, we have examined the electronic structure and transport properties of Yoshimori-type helimagnets, focusing on the impact of an external magnetic field on their band structure, non-reciprocal conductivity and photoconductivity. Unlike helimagnets stabilized by the Dzyaloshinskii-Moriya interaction, Yoshimori-type systems rely on competing exchange interactions, leading to spontaneously broken chiral symmetry. When a magnetic field is applied perpendicular to the spin rotation plane, the system undergoes a transition from a helimagnetic to a conical and eventually a ferromagnetic phase. Crucially, the conical phase exhibits band asymmetry, resulting in a finite second-order conductivity and non-zero injection photoconductivity. These findings highlight the potential of centrosymmetric helimagnets for chiral transport applications and motivate further exploration of their spintronic functionalities.

\textit{Acknowledgements}\textemdash The authors gratefully acknowledge the Research Infrastructures at the Center for Computational Materials Science at the Institute for Materials Research for allocations on the MASAMUNE-IMR supercomputer system (Project No. 202112-SCKXX-0510).  M.S.B. acknowledges support from Leverhulme Trust (Grant No. RPG-2023-253).
D.-A.D. had support from the Engineering and Physical Sciences Research Council Standard Research Studentship (DTP) EP/T517823/1.
A. A. acknowledges Wojciech J. Jankowski for helpful discussions as well as funding from the Cambridge International Scholarship awarded by the Cambridge Trust.  
R.-J.S. acknowledges funding from an EPSRC ERC underwrite grant  EP/X025829/1 and a Royal Society exchange grant IES/R1/221060 as well as Trinity College, Cambridge. R.V.B. and M.S.B. are grateful to E-IMR center at the Institute for Materials Research, Tohoku University, for continuous support.
% ****** End of file apssamp.tex ******

%\bibliography{bibliography.bib}

\clearpage % Ensure the appendix starts on a new page
\onecolumngrid % Switch to one-column mode (helps prevent layout issues)

\setcounter{equation}{0}
\setcounter{figure}{0}
\setcounter{table}{0}
\setcounter{page}{1}
\makeatletter
\renewcommand{\theequation}{S\arabic{equation}}
\renewcommand{\thefigure}{S\arabic{figure}}
\renewcommand{\thetable}{S\arabic{table}}
\renewcommand{\thesection}{S\arabic{section}}
\renewcommand{\thepage}{\arabic{page}}

\section{Electronic conductivity from Boltzmann transport}
In this section, we provide a short derivation of the $n$-th order electronic conductivity expression using Boltzmann transport formalism under the constant relaxation time approximation. Under this assumption, the evolution of the carrier distribution function $f$ under an external electric field $E$ in one dimension is given by
\begin{equation}
    -\frac{e}{\hbar}E\frac{\partial f}{\partial k_x} = -\frac{f-f_0}{\tau},
    \label{eq:boltzmann}
\end{equation}
where $\tau$ is the constant relaxation time, and $f_0=f_0(E_m,\mu,T)$ is the equilibrium Fermi-Dirac distribution function for a band $m$ at chemical potential $\mu$ and temperature $T$. Performing a series expansion in the electric field such that $f=f_0+f_1+f_2 + ...$, with $f_n \propto E^n$, one can easily find an expression for the $n$-th order correction to $f_0$ from Equation \ref{eq:boltzmann} as \citeB{ideue2017bulk-s}
\begin{equation}
    f_n=\Big(\frac{e \tau E}{\hbar}\Big)^n \frac{\partial^n f_0}{\partial k^n}.
    \label{eq:norder_distribution}
\end{equation}

The $n$-th order current density can be calculated as 
\begin{equation}
    j_x^{(n)} = -e\int \frac{\mathrm{d}k}{2\pi} \sum_m v_m(k) f_n(E_m,\mu,T) E^n,
\end{equation}
where $v_m(k)=-\frac{1}{\hbar}\frac{\partial E_m}{\partial k}$ is the velocity of band $m$, and the sum goes over all the bands. Using the fact that $j_x^{(n)}=\sigma^{(n)}E^n$ and the expression in Equation \ref{eq:norder_distribution}, we obtain the expression for the $n$-th order electronic conductivity used in the main text:
\begin{equation}
    \sigma^{(n)}(\mu,T) = -e \int \frac{\mathrm{d}k}{2\pi} \sum_m v_m(k) \Big(\frac{e\tau}{\hbar}\frac{\partial}{\partial k}\Big)^{n}\!f_0\big(E_m,\mu,T\big).
\end{equation}

\section{Quantum Geometry and optical responses}

We here expand on the quantum geometric approach used in the paper to calculate the injection and shift photoconductivities. Generally, in optical response theory, it is possible to express various photoconductivities derived from a perturbation theory of the electric dipole Hamiltonian \citeB{Sipe1993-s,Sipe2000-s} in terms of geometric quantities defined on a manifold of quantum states labeled by the wavevector $\textbf{k}$ \citeB{Ahn2020-s,bouhon2023quantum-s, Jankowski2025-s,Ahn2021-s}. The tangent vector $r_{mn}^a$ on such a manifold can be defined as 

\begin{equation} \label{eqn:tangent_vect}
    r_{mn}^a = i \delta_{mn} \partial_a + \bra{u_{m\kv}}  i \partial_a  \ket{u_{n\kv}},
\end{equation}

\noindent where $\partial_a = \frac{\partial}{\partial k^a}$, and $\ket{u_{n\kv}}$ represents the periodic part of the Bloch eigenstate in band $n$ and at wavevector $\textbf{k}$. Importantly, the second term in the equation above can be identified as the non-Abelian Berry connection $\textbf{A}_{mn}^a (\kv)$ \citeB{vanderbilt2018berry-s}.

In order to relate the above to optical responses, one can note that the transition dipole moment $\bra{\psi_{m\kv}}\hat{\textbf{r}}\ket{\psi_{n\kv}} = \textbf{r}_{mn}$. Then, introducing the transition dipole operator $\hat{e}^{mn}_a = r_{mn}^a \ket{u_{m\kv}} \bra{u_{n\kv}}$ ($m \neq n$), the quantum geometric tensor on this manifold can be defined following Ref. \citeB{Ahn2021-s} using the Hilbert-Schmidt inner product as 

\begin{equation}
    Q^{mn}_{ab} = \langle \hat{e}^{mn}_a, \hat{e}^{mn}_b \rangle = \text{Tr} \left(\hat{e}^{mn\dagger}_a \hat{e}^{mn}_b  \right) = r^a_{nm}r^b_{mn},
\end{equation}

\noindent where the second equality arises from the definition of the Hilbert-Schmidt inner product \citeB{Ahn2021-s}. Note that owing to its relationship with the transition dipole moment, the quantum geometric tensor enters the expression for both the shift and injection integrands. 

Since the injection current arises due to the differing band velocities of the valence and conduction bands, its conductivity in $d$ dimensions is given by \citeB{Ahn2021-s}

\begin{equation} \label{eqn:injection}
    \sigma^{c;ab}_{\text{i}} = -\frac{\pi e^3 \tau}{\hbar^2} \sum_{m,n} \int \frac{\dd^d k}{(2\pi)^d} \delta(\om - \om_{mn}) f_{nm} Q^{mn}_{ba} \partial_c \om_{mn},
\end{equation}

\noindent where $\tau$ is the mean free time for electrons in the material. The $c$ index labels the direction of the current, while the $a$ and $b$ indices indicate the polarization of the oscillating electric field.

On the contrary, the shift photocurrent requires the definition of a Hermitian connection on the aforementioned manifold (see Ref. \citeB{Ahn2021-s}),

\begin{equation}
    C^{mn}_{acb} = \langle \hat{e}^{mn}_a, \nabla_c \hat{e}^{mn}_b \rangle = r^a_{nm}r^{b;c}_{mn},
\end{equation}

\noindent where ${r^{b;c}_{mn} = \partial^{c} r^{b}_{mn} + i(A^{c}_{mm} - A^{c}_{nn} ) r^{b}_{mn}}$ is the generalized covariant derivative of the transition dipole moment, and as before, $\langle A,B \rangle$ denotes the Hilbert-Schmidt inner product. The shift conductivity is then given by \citeB{Ahn2021-s}

\begin{align}
\begin{split} \label{eqn:shift}
    \sigma^{c;ab}_{\text{s}} &= -\frac{\pi e^3}{2\hbar^2} \sum_{m,n} \int \frac{\dd^d k}{(2\pi)^d} \delta(\om - \om_{mn}) f_{nm} i\Big(C^{mn}_{bca}-C^{nm}_{acb}\Big) \\
    &= -\frac{\pi e^3}{2\hbar^2} \sum_{m,n} \int \frac{\dd^d k}{(2\pi)^d} \delta(\om - \om_{mn}) f_{nm}(R^{c;a}_{mn} - R^{c;b}_{nm}) Q^{mn}_{ba},
\end{split}
\end{align}

\noindent where $R^{c;a}_{mn} = i\partial^c {\ln r^{a}_{mn}} - (A^c_{mm}-A^c_{nn})$ is the shift vector as defined in the main text. As with the injection conductivity in equation \ref{eqn:injection}, the $c$ index labels the direction of the current, while the $a$ and $b$ indices denote the polarization of the oscillating electric field. The latter formulation in terms of the shift vector suggests that shift currents are caused due to the movement of the Wannier centers during the transition \citeB{Sipe2000-s,resta_shift-s} (indicated by the $A^c_{mm}-A^c_{nn}$ term \citeB{king1993theory-s}) and may therefore be related to the in-plane polarisation of the material \citeB{agarwal2024shift-s,Fregoso2017-s}.

Note that the expressions for the shift and injection photoconductivities given above are generalised forms of those in the main text, where the $1$-dimensionality of the system forces $a=b=c$. 

Finally, the above photoconductivities can be related to the experimentally measurable second-order DC photocurrents using \citeB{Ahn2020-s,Ahn2021-s}

\begin{equation}
    j^c_{\text{i/s}} = 2 \sum_{a,b} \sigma^{c,ab}_{\text{i/s}}(\om)\mathcal{E}^a(\om)\mathcal{E}^b(-\om),
\end{equation}

\noindent where the currents $j^c_{\text{i/s}}$ are induced by the oscillating electric fields $\mathcal{E}^{a,b}(\om)$.

\end{document}